\documentclass[12pt]{iopart}
\usepackage{iopams,graphicx}

\newcommand{\gr}{\mathbf{r}}
\newcommand{\gk}{\mathbf{k}}
\newcommand{\Psiop}{\hat{\Psi}}
\newcommand{\Psiopd}{\hat{\Psi}^\dagger}
\newcommand{\cl}{\hat{c}_l}
\newcommand{\cri}{\hat{c}_r}
\newcommand{\cs}{\hat{c}_s}
\newcommand{\ca}{\hat{c}_a}
\newcommand{\cld}{\hat{c}_l^\dagger}
\newcommand{\crid}{\hat{c}_r^\dagger}
\newcommand{\csd}{\hat{c}_s^\dagger}
\newcommand{\cad}{\hat{c}_a^\dagger}
\newcommand{\g}{\hat{g}}
\newcommand{\gd}{\hat{g}^\dagger}

\begin{document}

\title[Noise Thermometry with a Bosonic Josephson Junction]{A Primary Noise Thermometer for Ultracold Bose Gases}

\author{R Gati, J Esteve, B Hemmerling, T B Ottenstein, J Appmeier, A Weller, M K Oberthaler}

\address{Kirchhoff-Institut f\"ur Physik,
Universit\"at Heidelberg, Im Neuenheimer Feld 227, 69120 Heidelberg,
Germany} \ead{noisethermometry@matterwave.de}

\begin{abstract}
We discuss in detail the experimental investigation of thermally
induced fluctuations of the relative phase between two weakly
coupled Bose-Einstein condensates. In analogy to superconducting
Josephson junctions, the weak coupling originates from a tunneling
process through a potential barrier which is obtained by trapping
the condensates in an optical double-well potential. The observed
fluctuations of the relative phase are in quantitative agreement
with a many body two mode model at finite temperature. The agreement
demonstrates the possibility of using the phase fluctuation
measurements in a bosonic Josephson junction as a primary
thermometer. This new method allows for measuring temperatures far
below the critical temperature where standard methods based on time
of flight measurements fail. We employ this new thermometer to probe
the heat capacity of a degenerate Bose gas as a function of
temperature.
\end{abstract}

 \submitto{New J. Phys.
8 189 \quad doi:10.1088/1367-2630/8/9/189}

\maketitle



\section{Introduction}
Among other methods~\cite{Schumm:97,saba:05}, a stable double-well
potential for Bose-Einstein condensates can be realized by
superimposing a three-dimensional harmonic trapping potential and a
one-dimensional optical lattice with large periodicity
\cite{Albiez:05}. If the height of the potential barrier in the
center of the trap becomes comparable to the chemical potential, the
Bose-Einstein condensate (BEC) is split into two parts. If the
barrier height is not too high, the two split condensates are still
weakly coupled via tunneling through the barrier in analogy to
superconducting Josephson junctions \cite{Likharev:79,Barone:82} and
superfluid Helium weak links~\cite{Davis:02}.

At equilibrium, the relative phase between the two coupled
condensates fluctuates because of quantum fluctuations
\cite{Sols:94,Imamoglu:97,Leggett:01} and thermally induced
fluctuations \cite{Pitaevskii:01}. The experiments discussed in this
paper are performed in the so-called 'classical Josephson regime'
where quantum mechanical fluctuations are negligible and thermally
induced fluctuations can be treated classically. As shown in
\cite{Pitaevskii:01}, the phase fluctuations in this regime depend
only on a single parameter given by the ratio of the thermal energy
and the tunneling coupling strength between the two wells.

The dependence of the phase fluctuations on temperature allows for
the application of the fluctuation measurements for thermometry
\cite{Gati:06}. Hereby the coherence factor or to be more precise
the distribution of the relative phases, is measured experimentally
and compared to the theoretical prediction. Standard methods to
estimate temperature in Bose-Einstein condensed atomic samples
usually rely on time of flight measurements. The temperature is
deduced after ballistic expansion either from the expansion velocity
of the thermal cloud or from the ratio of the condensed to the
non-condensed populations. This method starts to fail for
temperatures far below the critical temperature where the number of
particles in the thermal cloud becomes small compared to the number
of particles in the BEC. However, even in this regime, the phase
fluctuation method can be applied as the coupling strength can be
tuned via the barrier height in order to make the bosonic Josephson
junction (BJJ) sensitive to thermally induced effects.

Experimentally the relative phase between the Bose-Einstein
condensates can be accessed by time of flight during which the two
condensates expand and interfere \cite{Andrews:97}. The resulting
density profile is analogous to the intensity profile observed with
coherent light in double-slit interference experiments and the
relative phase can be deduced from the position of the interference
peaks with respect to their envelope. Interference patterns are
observable for any temperature as far as the Bose-Einstein
condensate fraction is large enough for detection. Hereby the
visibility of the patterns increases with decreasing temperature
making the investigation of the thermally induced fluctuations
accessible for a wide range of temperatures.

Thermally induced phase fluctuations have been observed so far in
elongated Bose-Einstein condensates
\cite{Dettmer:01,richard:010405}. In these quasi 1D BEC the
coherence of the whole cloud is diminished by the thermally
populated low lying excitations. In time of flight these excitations
are revealed as density fringes on the BEC envelope. For low
temperatures the contrast of the fringes decreases and vanishes when
the typical length of the phase fluctuations becomes larger than the
longitudinal extent of the condensate. The situation in the
presented double-well potential is different. Here only the
coherence between the two wells is affected by thermal processes and
not the coherence within each well. The fluctuations decrease and
vanish for very low temperatures but increases for raising the
barrier height.

In this letter we would like to discuss the experiments presented in
\cite{Gati:06} in more detail and focus on the application of the
measurements for thermometry. We justify the classical model
introduced before to describe the fluctuations of the relative phase
by comparing it with the prediction of a many body two mode
approach. By using a more advanced calibration methods we improved
the determination of the tunneling coupling leading to a more
quantitative agreement of the experimental findings with the
theoretical prediction. This allows for the application of the phase
fluctuation measurements for primary thermometry, i.e. the
thermometer is calibrated directly by the theoretical model and thus
the calibration with temperature measurements using other methods is
not necessary.

\section{Phase fluctuations within the two mode approximation}
In order to use the phase fluctuation measurement as a tool for
thermometry, a precise theoretical model is needed to convert the
measured fluctuations into a temperature. In this section, we
present in detail the two mode model that we use for this purpose.
In this model, the condensate particles can occupy only two single
particle states while being in thermal equilibrium with a bath
composed of the non condensed atoms. We will first shortly introduce
the two mode model which has attracted tremendous interest in the
literature~(see references 2 to 20 in \cite{Ananikian:06}). We will
then show that, because of its relative simplicity, this model
allows exact numerical calculations of the phase fluctuations (or
any other quantity) at finite temperature. The drastic reduction of
the Hilbert space dimension due to the two mode approximation makes
the two mode model one of the few exactly solvable many body systems
(see for example~\cite{Milburn:97,Anglin:01,dunningham:033601}). We
will compare our numerical results with analytical results that can
be obtained in the low temperature and the high temperature limit.

\subsection{The two mode approximation}
Starting from the general many body problem of a gas of interacting
bosons, the two mode approximation consists of restricting the
available single particle states to two states $|\phi_1\rangle$ and
$|\phi_2\rangle$. Fixing the total atom number in the sample to a
given value $N$, a basis of the system can then be obtained by
considering the following set of Fock states
\begin{equation}\label{eq.basis}
    |n\rangle = |N-n : \phi_1, n : \phi_2 \rangle \ \ {\rm where}
    \ \ n=0,1, \ldots, N \, .
\end{equation}
The dimension of the Hilbert space is then reduced to $N+1$. This
drastic reduction of the dimension allows exact numerical quantum
calculations for the atom number range accessed in our experiments
($N < 10^4$).

The double-well trap used in our experiment naturally provides a
geometry where the two mode approximation can be applied. Indeed, a
Bogoliubov calculation of the excitation spectrum or a calculation
of the Gross-Pitaevskii eigenenergies shows that the first
excitation lies close to the ground state while the second
excitation is well above. As in the non-interacting case, such a
situation occurs only if the barrier separating the two condensates
is sufficiently high (higher than the chemical potential in each
well). Supposing this condition to be fulfilled, we choose for the
two single particle states $|\phi_1\rangle$ and $|\phi_2\rangle$,
the first two eigenstates of the Gross-Pitaevskii equation
$|\phi_s\rangle$ and $|\phi_a\rangle$: $|\phi_s\rangle$ being the
symmetric ground state and $|\phi_a\rangle$ the antisymmetric first
excited state\footnote{We assume the wavefunctions $\phi_s$ and
$\phi_a$ to be normalized to one.}. The corresponding eigenvalues
are given by
\begin{equation}
\mu_{s,a}=\int{\phi_{s,a}^* \left(-\frac{\hbar^2}{2m}\nabla^2 + V +
gN|\phi_{s,a}|^2 \right)\phi_{s,a} \, d\gr} \, .
\end{equation}
where $g$ is the effective interaction strength in 3D.

Introducing the creation operators $\csd$ and $\cad$ associated with
the two wavefunctions $\phi_s$ and $\phi_a$, the field operator
writes in the two mode approximation
\begin{equation}\label{eq.2mode}
    \Psiop = \cs \phi_s + \ca \phi_a \, .
\end{equation}
Starting from the many body Hamiltonian
\begin{equation}\label{eq.htotal}
    \hat H = \int \left( \Psiopd \left(-\frac{\hbar^2}{2m} \nabla^2 + V \right) \Psiop +\frac{g}{2}
    \Psiopd \Psiopd \Psiop \Psiop \right)  d\gr \, ,
\end{equation}
and ignoring constant terms (terms proportional to the total atom
number), we obtain the following two mode
Hamiltonian~\cite{Ananikian:06,Spekkens:99}
\begin{eqnarray}
\hat{H}_{\rm 2M} & = &  - \frac{E_J}{N} (\csd \cs - \cad \ca) + \frac{E_C}{8} (\csd \ca +\cad \cs)^2 + \frac{\delta E_C}{4} (\csd \cs - \cad \ca)^2  \label{eq.hsa} \\
    & = & - \frac{E_J}{N} (\cld \cri + \crid \cl) + \frac{E_C}{8} (\crid \cri - \cld \cl)^2 + \frac{\delta E_C}{4} (\cld \cri + \crid \cl)^2 \label{eq.hlr}.
\end{eqnarray}
The last equality expresses the Hamiltonian in terms of creation and
annihilation operators in the left/right basis which we define as
$\cld = (\csd+\cad)/\sqrt{2}$ and $\crid = (\csd - \cad)/\sqrt{2}$.
The Josephson energy $E_J$, the charging energy $E_C$ and the
correction term $\delta E_C$ are defined as follow
\begin{eqnarray}
\kappa_{i,j} & = &  \frac{g}{2} \int |\phi_i|^2 |\phi_j|^2 d\gr \quad \quad (\textrm{with} \quad i,j=s,a) \\
E_C & = & 8 \kappa_{s,a} \label{eq.ec} \\
E_J & = & \frac{N}{2}(\mu_a - \mu_s) - \frac{N(N+1)}{2} (\kappa_{a,a} - \kappa_{s,s}) \label{eq.ej} \\
\delta E_C &  = & \frac{\kappa_{s,s} + \kappa_{a,a} - 2
\kappa_{s,a}}{4} \, .
\end{eqnarray}
The first term of the two mode Hamiltonian which is proportional to
$E_J$ describes the tunneling between the two wells. It is diagonal
in the symmetric/antisymmetric basis. The second term proportional
to $E_C$ describes the on-site interaction in each well and is
diagonal in the left/right basis. The last term is a correction term
that, as we will see, can usually be neglected.

In the high barrier limit, the different interaction coefficients
$\kappa_{i,j}$ can be considered to be equal and the correction term
$\delta E_C$ vanishes. In our experimental parameter range, we find
this term to be less than $10^{-5} \cdot E_J/N$ and less than
$10^{-3} \cdot E_C/8$. The correction term is included in our
numerical calculations, however its effect is so small that we will
neglect it in the following discussion. To obtain the values of
these three coefficients, we numerically solve the 3D Gross
Pitaevskii equation in our potential using a standard split-step
iteration algorithm and propagation in imaginary time. The precision
of the calculated values directly relies on the numerical precision
of the employed method. In particular, the Josephson energy requires
special care, because it critically depends on the difference of two
energies $\mu_s$ and $\mu_a$ which are almost degenerate.

\subsubsection{Eigenenergy spectrum}
\begin{figure}[htbp]
\centering\includegraphics[totalheight=10cm]{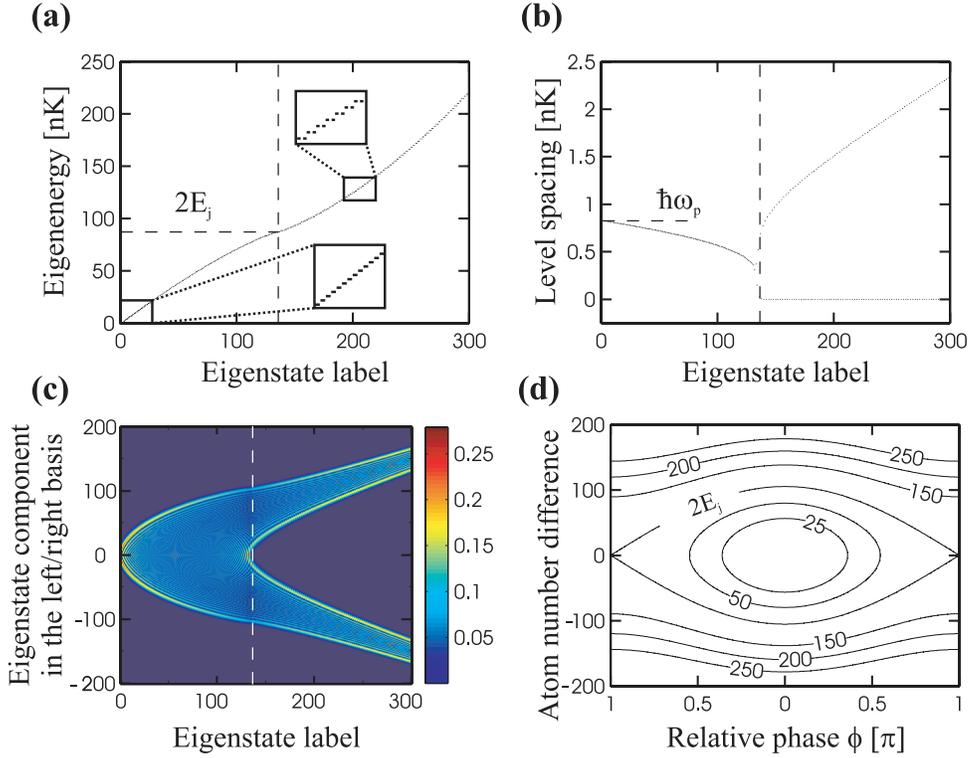}
\caption{Lower energy eigenvalues and eigenstates of the two mode
Hamiltonian in the Josephson regime ($E_C = 0.016$~nK, $E_J$=43~nK
and 3000 atoms): (a) eigenenergies, (b) eigenenergy differences, (c)
eigenstates components in the left/right Fock state basis $|N/2-n
:{\rm left},N/2+n:{\rm right}\rangle$, (d) Phase portrait of the
classical pendulum associated with the two mode model for different
total energies. At low energy $(E<2E_J)$, the linear part of the
spectrum corresponds to an oscillatory motion in the pendulum
picture and the associated eigenstates are delocalized in the
left/right basis. At higher energy, each level is two times
degenerate, which corresponds to the two possible rotation
directions of the pendulum twirling motion.} \label{fig.theo1}
\end{figure}

As mentioned above, the limited size of the Hilbert space allows
exact numerical diagonalization of the two mode Hamiltonian. Our
experimental parameter range is such that we always lie in the
Josephson regime where the two conditions $E_C \ll E_J$ and $E_C\gg
E_J/N^2$ are fulfilled. In this regime, a typical energy spectrum of
the many body system looks like the one shown in
Fig.~\ref{fig.theo1}. (a). At low energy ($E<2E_J$), the spectrum is
almost linear and the level spacing is approximately given by the
plasma frequency (see Fig.~\ref{fig.theo1} (b))
\begin{equation}
\omega_p = \frac{1}{\hbar} \sqrt{E_J \left(E_C + \frac{4
E_J}{N^2}\right)} \quad . \label{eq.plasma_freq}
\end{equation}
The corresponding eigenstates are delocalized over the two wells
(see Fig.~\ref{fig.theo1} (c)). In a classical picture, the two mode
Hamiltonian is equivalent to the Hamiltonian of a pendulum, with the
relative phase corresponding to the tilt angle and the population
imbalance to the momentum. In this picture these states correspond
to an oscillatory motion of the pendulum around its equilibrium (see
Fig.~\ref{fig.theo1} (d)). At higher energy ($E>2 E_J$), the
eigenenergies are grouped two by two. Each doublet of almost
degenerate eigenstates consists of a symmetric and an antisymmetric
cat-state (see Fig.~\ref{fig.theo1} (c)), however any asymmetry
between the two wells will localize the modes on the left or on the
right. In the classical picture, these states correspond to a
twirling motion of the pendulum. Here the pendulum has enough energy
to reach the top position and continues rotating in one direction.
The degeneracy corresponds to the two possible directions of
rotation. The energy of these states is dominated by the charging
energy term and thus increases quadratically with the eigenstate
label as expected for a free particle motion.

\subsection{Calculation of the coherence factor}
\label{sec_calc_coh}
As we will show now, our measurement of the phase fluctuations is a
direct measurement of the coherence of the system. The coherence of
a many body system can be quantitatively measured by considering the
first order spatial coherence function $g^{(1)}(\gr,\gr')$. In the
two mode approximation, it is easy to see from (Eq.~\ref{eq.2mode})
that the first order spatial coherence function does not depend on
the difference $(\gr-\gr')$ and is constant~\cite{Spekkens:99}. We
can then unambiguously define the coherence of the system $\alpha$
as the uniform value of the spatial coherence function.
\begin{eqnarray}\label{eq.g1}
    \alpha = g^{(1)}(\gr,\gr') & = \frac{\langle \Psiopd(\gr') \Psiop(\gr) \rangle}{\sqrt{\langle \Psiopd(\gr) \Psiop(\gr) \rangle \langle \Psiopd(\gr') \Psiop(\gr')
    \rangle}} \nonumber \\
    & = \frac{\langle \cld \cri + \crid \cl \rangle}{N} = \frac{\langle \csd \cs - \cad \ca
    \rangle}{N} \, .
\end{eqnarray}
In the two mode approximation, the coherence of the system is simply
given by the relative population difference between the symmetric
and the antisymmetric state.

A more descriptive measure for phase fluctuations is the mean fringe
visibility. The visibility of a single interference pattern is
always high and only reduced by an initial population imbalance
between the wells and the resolution of the detection method, but it
is independent on temperature \cite{Castin:97}. However, the mean
fringe visibility is the visibility of ensemble averaged
interference patterns and if the phase fluctuates from measurement
to measurement the averaging leads to a reduction of the signal
corresponding to a lower visibility. The equivalence between the
coherence factor defined above and the mean fringe visibility can be
established in the following way. In the two mode approximation, the
Fourier component of the field operator is $\Psiop(\gk) = \cs
\phi_s(\gk) + \ca \phi_a(\gk)$. The Fourier transform of the
symmetric and the antisymmetric state correspond to fringe patterns
with respectively a zero (bright central fringe) and a $\pi$ phase
(dark central fringe) modulated by an envelope $f(\gk)$. The
ensemble averaged mean atom velocity distribution can thus be
written as
\begin{eqnarray}
    n(\gk)  & = & \langle \Psiopd(\gk) \Psiop(\gk) \rangle \nonumber \\
            & = & n_s |\phi_s(\gk)|^2 + n_a |\phi_a(\gk)|^2 \nonumber \\
            & = & \left[ 1 + \frac{n_s-n_a}{N} \cos(k_x d) \right] f(\gk) =  \left[ 1 + \alpha \cos(k_x d) \right]
            f(\gk) \, ,
            \label{eq.contrast}
\end{eqnarray}
where $\alpha$ is the visibility of the interference patterns.
Assuming that the interactions do not significantly modify the
velocity distribution of the atoms after a time of flight,
Eq.~\ref{eq.contrast} shows that the mean visibility factor is equal
to the relative population difference between the symmetric and
antisymmetric states and thus to the coherence factor $\alpha$
(Eq.~\ref{eq.g1}).

\subsection{Finite temperature}
At finite temperature the coherence of the BJJ is determined by
quantum mechanical but also thermally induced fluctuations. To model
thermal effects we consider the two mode approximation for the
condensed fraction and treat the particles in higher lying excited
states as a thermal bath. We neglect the influence of the thermal
distribution on the ground and the first excited states and thus on
the parameters $E_C$ and $E_J$. This assumption is justified if the
density of the thermal cloud is low compared to the total peak
density.

This allows us to calculate the density matrix using the calculated
eigenenergy spectrum and thus any thermally averaged quantity. In
the following we compare the numerically obtained coherence factor
with analytic predictions for the low and the high temperature
limit.

\subsubsection{Low temperature limit} We define the low temperature
regime as the range corresponding to temperatures on the order of
the plasma frequency up to the Josephson energy. In this regime,
both quantum and thermal fluctuations play a role, but their overall
contribution is small enough so that the coherence factor is close
to 1. Following a Bogoliubov approach, we can calculate analytically
the expression of the coherence factor. As shown
in~\cite{Paraoanu:01}, starting from the two mode Hamiltonian, the
Bogoliubov transformation is straightforward since only one
quasiparticle mode can exist. The transformed Bogoliubov Hamiltonian
writes
\begin{equation}\label{eq.HBg}
    \hat{H}_{\rm Bg} = E_{\rm Bg}(N) + \hbar \omega_p (\gd \g + 1/2
    ) \, .
\end{equation}
$E_{\rm Bg}(N)$ is a constant energy term. The excitation creation
operator is $\gd = u \cad + v \ca$ where $u= \cosh \chi $, $v=\sinh
\chi$ and $\tanh 2\chi = E_C/(E_C+8 E_J /N^2)$~\cite{Paraoanu:01}.
The obtained excitation spectrum corresponds to the linear part of
the exact spectrum plotted in Fig.~\ref{fig.theo1} (a).

To determine the coherence factor, we calculate the number of atoms
in the antisymmetric state by
\begin{eqnarray}
    n_a & = & \langle \cad \ca \rangle \nonumber \\
       & = & \langle \gd \g \rangle (u^2+v^2) + v^2 \, .
\end{eqnarray}
We find the usual formula for the condensate depletion: the first
term corresponds to thermal fluctuations and the second term to
quantum depletion. The coherence factor is then
\begin{equation}\label{eq.cohBg}
    \langle \alpha \rangle = 1-2\frac{n_a}{N} \simeq
    1 -  \frac{1}{2}\sqrt{\frac{E_C}{E_J}}
    \left( \frac{1}{\exp(\beta \hbar \omega_p) -1} + \frac{1}{2}
    \right) \, .
\end{equation}
with $\beta=(k_BT)^{-1}$. In the last equality, we have used the
assumption that we are in the Josephson regime ($E_J/N^2 \ll E_C$).
The decoherence due to quantum fluctuations is proportional to
$\sqrt{E_C/E_J}$ and is always small in the Josephson regime. In
Fig.~\ref{fig.theo2}, we compare this analytical result with an
exact numerical calculation of $\alpha$ for different temperatures.
We observe that the Bogoliubov result is only verified when the
condensate depletion is sufficiently small which corresponds to
temperatures below the Josephson energy.

\subsubsection{High temperature limit} For temperatures much higher
than the mean level spacing, which is approximately given at low
energy by the plasma energy, a semi-classical calculation of the
thermal average fringe visibility is valid. The two mode Hamiltonian
corresponds to the following classical Hamiltonian
\cite{Raghavan:99}
\begin{equation}
H_{\rm Cl} = E_C \frac{n^2}{2} - E_J \sqrt{1-\frac{4 n^2}{N^2}} \cos
\phi \, ,
\end{equation}
where $n$ is half the atom number difference between the two wells
and $\phi$ is the relative phase between the two wells.

\begin{figure}[htbp]
\centering\includegraphics[totalheight=6cm]{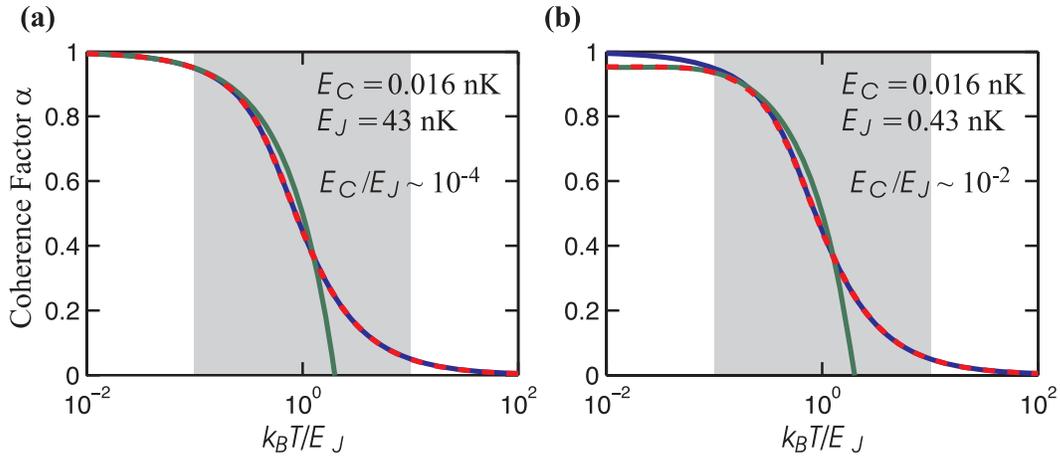}
\caption{Coherence factor $\alpha$ as a function of the ratio
$k_BT/E_J$ for 3000 atoms. The dashed red curve shows the result of
the exact numerical calculation. The green curve is the Bogoliubov
calculation (\ref{eq.cohBg}) which is valid only at low temperature
and correctly accounts for quantum fluctuations (right graph). The
blue curve is the result of the semi-classical calculation
(\ref{eq.classical}) which is valid at high temperature and also at
low temperature if quantum fluctuations are negligible (left graph).
The grey shaded area corresponds to the region where the measurement
of the coherence factor can be used for thermometry.}
\label{fig.theo2}
\end{figure}

In this picture, the coherence factor corresponds to the mean value
of $\cos \phi$ which writes
\begin{equation}
    \langle \cos \phi \rangle = \frac{1}{Z} \int \cos \phi \exp(-\beta H_{\rm Cl}) \ \ {\rm where} \ \ Z=\int \exp(-\beta
    H_{\rm Cl}) \, .
\end{equation}
In the Josephson regime, the relatively strong interaction term
prevents any contribution of large values of $n$ in the previous
integral. The coherence factor can then be approximated by
(\cite{Pitaevskii:01})
\begin{equation}\label{eq.classical}
    \alpha = \langle \cos \phi \rangle = \frac{\int \cos \phi \exp(-\beta E_J \cos \phi)}{\int \exp(-\beta E_J \cos \phi)} = \frac{I_1(\beta E_J)}{I_0(\beta
    E_J)} \, ,
\end{equation}
where $I_i(j)$ are the modified Bessel functions of the first kind.
Fig.~\ref{fig.theo2} shows the comparison of this analytical
expression with the exact numerical calculation for different
temperatures and two different Josephson energies. At high
temperature, the agreement is always very good. At low temperature
and if the Josephson energy is smaller than approximately $10 E_C$,
a small discrepancy can appear because of quantum fluctuations (see
right graph in Fig.~\ref{fig.theo2}). To use the coherence factor
measurement for thermometry, we always place ourselves in the
situation shown in the left graph of Fig.~\ref{fig.theo2} where the
agreement between the exact calculation and the semi-classical
formula (Eq.~\ref{eq.classical}) is better than $1\%$ . In this
case, the coherence factor only depends on the dimensionless
parameter $\beta E_J$, the conversion between the two quantities
being precisely given by Eq.~\ref{eq.classical}. Since the phase
fluctuations depend strongly on the temperature in the range of $0.1
E_J$ and $10 E_J$ (see grey shaded area in Fig.~\ref{fig.theo2}),
this regime is ideally suited for deducing the temperature from
phase fluctuation measurements. This dynamic range can furthermore
be tuned by adjusting the coupling strength $E_J$. Once the
parameter $\beta E_J$ is known from the coherence measurement, the
conversion to a temperature relies on a precise calculation of
$E_J$. The main uncertainty in this calculation comes from the atom
number uncertainty.

\section{Experimental protocol}
In this section, we give an overview of our experimental thermometry
procedure. Our experimental setup has already been described in
detail in~\cite{Gati:061}.
\begin{figure}[h!]
\centering\includegraphics[totalheight=5cm]{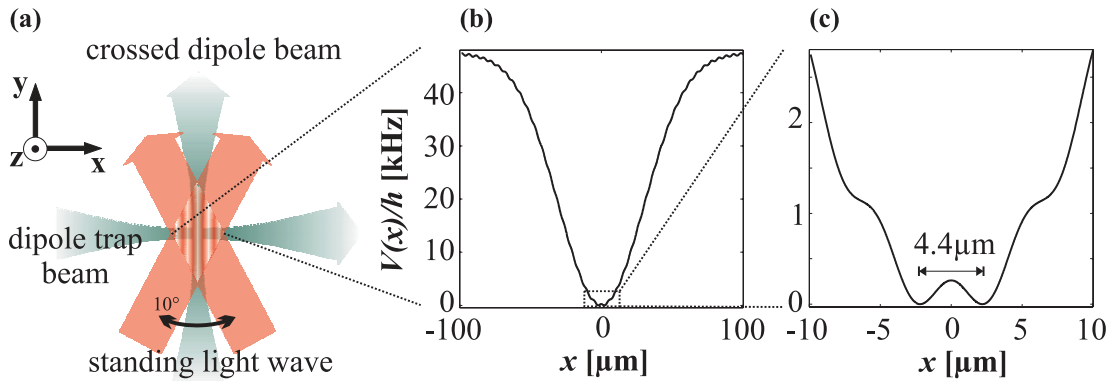}
\caption{\label{fig.setup} Experimental setup and realization of the
double-well potential by the superposition of a harmonic trap and an
optical lattice with large periodicity. (a) is a sketch of the laser
beams generating the optical potentials. Two orthogonal dipole trap
beams at 1064~nm (grey) create a 3D harmonic confinement and two
laser beams at 830~nm crossing under an angle of $10^\circ$ generate
the optical lattice (red) with a periodicity of
$\lambda=4.8$~$\mu$m. (b) shows the effective potential resulting
from the superposition of the dipole trap and the optical lattice on
the scale of the Gaussian dipole trap beam. (c) is the potential in
the center of the trap revealing the effective double-well trap with
a separation of the two wells of $4.4~\mu$m.}
\end{figure}
The experiment is performed as follows. We first generate a single
Bose-Einstein condensate by loading a precooled $^{87}$Rb sample in
a 3D optical dipole trap as indicated in Fig.~\ref{fig.setup} (a).
The harmonic trapping frequencies are $\omega_x=2\pi \cdot 90(2)$~Hz
and $\omega_{y,z}=2\pi \cdot 100(2)$~Hz. Evaporative cooling is
performed until the number of particles in the trap is lowered
usually to about 2000 to 4000 and the lowest accessible temperature
$T=15(4)$~nK is reached. Holding the atoms in the trap for a given
time allows for increasing the temperature of the sample in a
controlled way. The heating rate in the optical trap is about 2~nK/s
for a classical gas (see Sec.~\ref{sec_thermometry}). Once the final
temperature is reached, the harmonic potential is transformed into
the double-well potential by slowly ramping up a standing light wave
of the form $V=V_0/2\cdot(1+\cos(2\pi\cdot x/\lambda))$ with
$\lambda=4.8(2)~\mu$m. The resulting potential is shown in
Fig.~\ref{fig.setup} (b) and (c). The optical lattice is generated
by two interfering laser beams (830~nm) crossing under an angle of
$10^\circ$. The angle is chosen such that the distance between the
two wells is larger than the optical resolution of $3.2(2)~\mu$m
(sparrow-criterion). The barrier is ramped up in 300~ms in order to
prevent excitation of the condensates in each well and to guarantee
thermal equilibrium. The coupling strength due to tunneling between
the two wells can be adjusted by changing the barrier height.
Typically, we vary $V_0/h$ between 500~Hz and 2500~Hz (the chemical
potential ranges between 600~Hz and 900~Hz). Under these conditions,
the distance between the two wells is approximately $4.4~\mu$m as
shown in Fig.~\ref{fig.setup} (c).

\begin{figure}[h!]
\centering\includegraphics[totalheight=6cm]{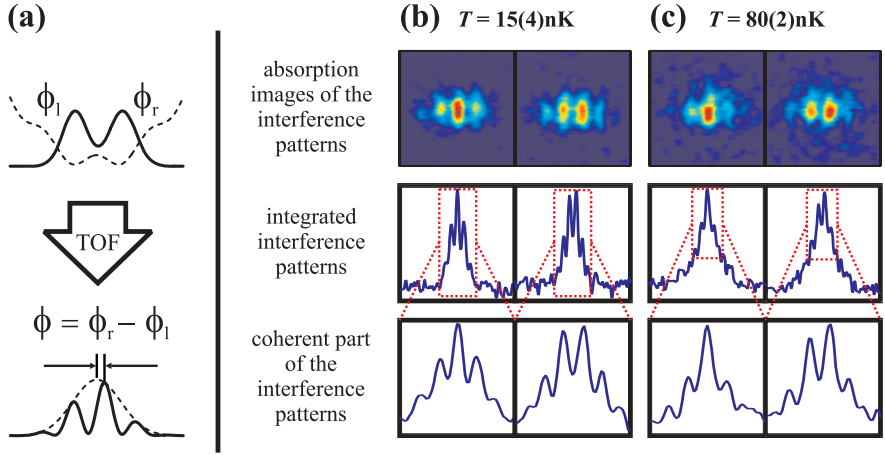}
\caption{\label{fig.interference} Matter-wave interference patterns.
(a) is a sketch of the interference experiments. Once the
double-well trap is turned off, the matter-wave packets expand,
overlap and interfere. (b) corresponds to typical interference
patterns for low temperature and (c) to typical interference
patterns for high temperature. The integrated patterns show a clear
interference signal for all temperatures (central graphs). However,
for high temperature also a broad background is visible,
corresponding to the distribution of the thermal atoms after the
expansion time. In order to find the coherent interference patterns,
this background is subtracted (lower graphs).}
\end{figure}

The relative phase between the two condensates is measured after a
time of flight of 5 or 6~ms. Typical interference patterns are shown
in Fig.\ref{fig.interference} (b) and (c).
Fig.\ref{fig.interference} (b) corresponds to measurements at low
temperature ($T=15(4)$~nK) and (c) to high temperature
($T=80(2)$~nK). The single images show a clear interference signal
with high visibility (usually larger than $50\%$). In order to fit
the relative phase, the images are integrated transversally to the
interference patterns (central graphs). At high temperature a broad
incoherent background becomes visible resulting from the ballistic
expansion of the thermal atoms. To fit the interference patterns at
high temperature, this background is subtracted. The resulting
profiles (lower graphs) thus correspond only to the coherent part of
the matter wave interference. The fitting procedure leads to an
uncertainty of the relative phase of about $0.13\pi$.

\begin{figure}[h!]
\centering\includegraphics[totalheight=3.5cm]{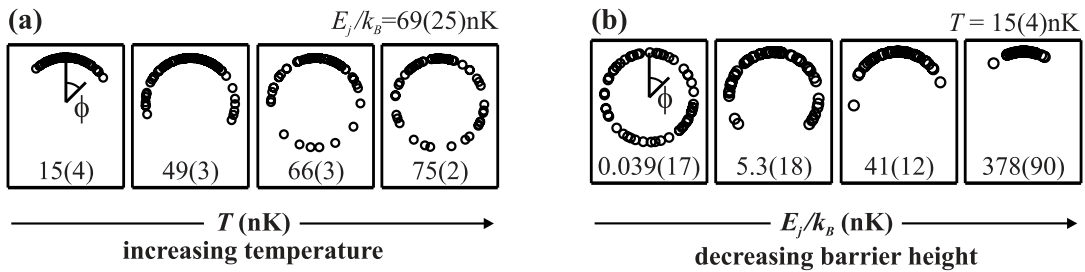}
\caption{\label{fig.phase_fluctuations} Qualitative behavior of the
fluctuations of the relative phase. The black circles correspond to
single measurements of interference patterns in a polar plot. (a)
shows the change of the fluctuations as a function of temperature
where the coupling strength is kept constant. With increasing
temperature the fluctuations increase. (b) shows the behavior of the
fluctuations for decreasing barrier heights leading to an increase
of the coupling strength. The temperature is kept constant. Due to
the increase of the coupling the fluctuations are decreased
demonstrating the stabilizing effect of the tunneling coupling.}
\end{figure}

Repeating the interference experiments reveals that the relative
phase is not constant but fluctuates around zero as shown in
Fig.~\ref{fig.phase_fluctuations}. The fluctuations increase with
increasing temperature and decrease with increasing tunneling
coupling. This shows qualitatively that two competing processes are
relevant for the thermally induced fluctuations of the relative
phase, namely thermal effects leading to a randomization of the
phase and the coherent tunneling coupling to a stabilization of the
phase.

\section{Coherence factor - scaling law for phase fluctuations}
\label{sec_coherence} 
\begin{figure}[h!]
\centering\includegraphics[totalheight=6.5cm]{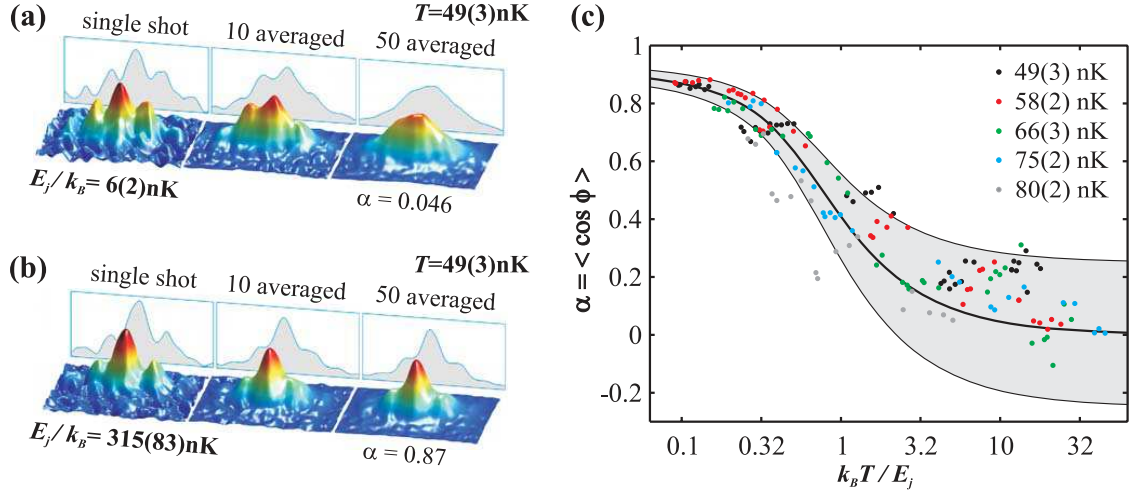}
\caption{\label{fig.coherence} Experimental investigation of the
coherence factor $\alpha$. (a) shows the transition from coherent
single realizations to incoherent ensemble averages. The first
picture shows that for single realizations a clear interference
picture is observed. For small tunneling coupling ($E_J=6(2)$~nK)
compared to the thermal energy ($T=49$~nK) the averaging over many
images leads to the loss of the visibility which is directly
connected to the loss of the coherence ($\alpha=0.046$). (b)
visualizes that for large coupling ($E_J=315(83)$~nK) the averaging
over many realizations does not destroy the coherence
($\alpha=0.87$) and leads only to a small degradation of the
visibility. (c) shows the quantitative behavior of the coherence
factor $\alpha$ as a function of the scaling parameter $k_BT/E_J$.
Each data point corresponds to the averaging over at least 28 (in
average 40) single measurements for different $T$ and $E_J$. The
temperature is measured independently with time of flight
measurements and the coupling strength is deduced from 3D
simulations of the BJJ using independently measured system
parameters (potential parameters and atom numbers). The experimental
error of $k_BT/E_J$ is about $\pm 30\%$. The central black line
corresponds to the prediction of the classical theory and also takes
the uncertainty of the phase-fitting into account. The grey shaded
area shows twice the expected standard deviation of the coherence
factor due to the finite number of measurements. The behavior of the
coherence factor is confirmed over a three orders of magnitude
change of the scaling parameter.}
\end{figure}

The coherence factor $\alpha$
(Eq.~\ref{eq.classical}) is in the thermodynamic limit a measure for
the fluctuations of the relative phase \cite{Gati:06} and is
directly connected to the visibility of the ensemble averaged
interference patterns as discussed in Sec.~\ref{sec_calc_coh}. It
predicts the coherence factor of the BJJ if quantum mechanical
fluctuations are negligible ($\hbar \omega_p \ll k_BT$) and thus
corresponds to an upper bound for the measurements. For large
$k_BT/E_J$ the phase fluctuates strongly from shot to shot and
causes a reduction of the ensemble averaged visibility. This
corresponds to a small coherence factor as visualized in
Fig.~\ref{fig.coherence} (a). By increasing the tunneling coupling
the coherence of the system can be regained as shown in
Fig.~\ref{fig.coherence} (b).

The quantitative measurement of the coherence factor $\alpha$ as a
function of the scaling parameter $k_BT/E_J$ is shown in
Fig.~\ref{fig.coherence} (c). For this, we performed up to 100
measurements with a condensate fraction ranging between 1000 and
4000 atoms at different temperature and different barrier heights
(which means that at high temperatures the total number of atoms is
approximately 10000). In Fig.~\ref{fig.coherence} (c) every data
point corresponds to a subset extracted from these measurements for
different atom number ranges (e.g. 2000 to 2500, 2500 to 3000
etc.~atoms in the condensate fraction). The Josephson energy $E_J$
is then calculated for every point at the given trap parameters and
the mean atom numbers in the BEC by numerically solving the Gross
Pitaevskii equation in 3D. The influence of the thermal atoms is
neglected. Each data point represents at least 28 single
measurements and on average 40 measurements. The coherence factor is
calculated by averaging over the cosine of the fitted phases.

The temperature corresponding to the data points is measured with
three different methods. In the single interference images it is
possible to fit the thermal background transversally to the
interference patterns. The temperature is then connected to the
waist of the thermal cloud and to the time of flight. Additionally,
independent time of flight measurements are performed releasing the
atoms from the harmonic trap and the temperature deduced from both
the expansion velocity of the thermal cloud and the ratio between
condensate and thermal fraction. All three methods lead within the
experimental error to the same results. The temperatures used for
$k_BT/E_J$ in Fig.~\ref{fig.coherence} (c) are extracted from the
time of flight measurements using the last method.

The typical error of $k_BT/E_J$ is $\pm 30\%$. The error in $E_J$
results from the uncertainty of the atom numbers, the trapping
frequencies, the barrier height and the lattice spacing of the
periodic potential. The error in $T$ results from the fitting error
of the waists and amplitudes of the double gaussian distribution of
the independent time of flight measurements.

The central black line in Fig.~\ref{fig.coherence} (c) shows the
theoretical prediction for the coherence factor taking the fitting
error of the relative phases into account. The influence of the
fitting error on the coherence factor can be estimated by
introducing an additional fluctuating phase and averaging over it.
The distribution of the additional phase is approximated by a box
function with a standard deviation corresponding to the fitting
error. The averaging leads to a reduction of the coherence factor
\begin{eqnarray}
\alpha'  = \langle \cos(\phi) \rangle' & = \frac{
\int_{-\phi_0}^{\phi_0} d\phi' \int_{-\pi}^{\pi} d\phi \cos(\phi -
\phi') \exp(E_J/k_BT \cos(\phi))}{\int_{-\phi_0}^{\phi_0} d\phi'
\int_{-\pi}^{\pi}
d\phi \exp(E_J/k_BT \cos(\phi))} \nonumber \\
& = \frac{\sin(\phi_0)}{\phi_0} \frac{\int_{-\pi}^{\pi} d\phi
\cos(\phi) \exp(E_J/k_BT \cos(\phi))}{\int_{-\pi}^{\pi} d\phi
\exp(E_J/k_BT \cos(\phi))} \nonumber \\
& = \frac{\sin(\phi_0)}{\phi_0} \cdot \alpha \, .
\end{eqnarray}
For our experiments $\phi_0$ is equal to $0.23\pi$ which corresponds
to a fitting error of
$\sqrt{\frac{1}{2\phi_0}\int_{-\phi_0}^{\phi_0}{\phi^2 d\phi}} =
0.13\pi$ of the relative phase and to a reduction of the coherence
factor $\alpha' = 0.92 \alpha$. The grey shaded area shows twice the
standard deviation of the coherence factor resulting from the finite
number of measurements. For about 40 measurements it can be
estimated by $\Delta \alpha \approx (1-\alpha)/8$.

The observed behavior of the coherence factor is consistent with the
two mode model prediction over three orders of magnitude of the
scaling parameter $k_BT/E_J$. For small values of $k_BT/E_J$ the
coherence factor is in close agreement with the theoretical
prediction. However, for $k_BT/E_J>2$ the data points lie within the
experimental error but are mainly localized above the curve. This
deviation can be explained by the fact that the BJJ is not
thermalized for small $E_J$ (see Sec.~\ref{sec_thermometry}). The
points corresponding to high temperature lie outside the shaded
region showing a lower degree of coherence. A disagreement in this
regime can also be expected as the temperature of 80~nK is close to
the critical temperature of $T_c\approx87$~nK. In this regime, the
excitation spectrum may not be correctly accounted for by the two
mode model.

\section{Thermalization} \label{sec.thermalization} The theoretical
prediction for the coherence factor is only valid in thermal
equilibrium. In order to check for thermal equilibration, different
experimental tests have been performed.
\begin{figure}[h!]
\centering\includegraphics[totalheight=6.5cm]{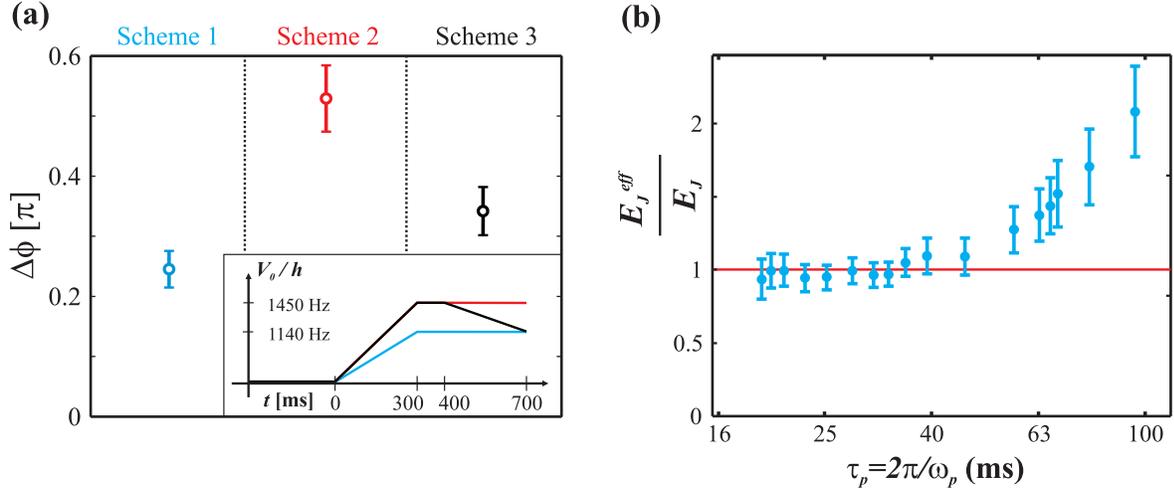}
\caption{\label{fig.thermalization} Experimental test for thermal
equilibration of the BJJ. In (a) three different ramps as indicated
are compared. The decrease of the fluctuations for scheme 3 with
respect to scheme 2 indicates the thermalization process and the
stabilizing effect of the tunneling coupling. (b) shows the
comparison of the measured coherence factors and the theoretical
expectation. The experimental $E_{j}^{eff}$ are deduced by fitting
the measured coherence factor to the theoretical prediction. For
tunneling times $\tau_p\leq50$~ms the ratio is 1 and no correction
is needed. For $\tau_p>50$~ms the ratio increases as the BJJ is not
completely thermalized. Thermalization takes only place if
$\tau_p\ll t_{ramp}=300$~ms, which is the time for ramping up the
potential barrier.}
\end{figure}

We have compared the fluctuations of the relative phase after three
different ramping schemes as shown in Fig.~\ref{fig.thermalization}
(a). For these measurements the atom number is 4000(400). In scheme
1, the barrier is ramped up to a low value ($V_0=1140$~Hz) within
300~ms and then kept the barrier constant for 400~ms. Scheme 2 is
similar to scheme 1 but with a higher final value of the barrier
($V_0=1450$~Hz). As expected, the fluctuations are small for scheme
1 and larger for scheme 2. Scheme 3 is a combination of the two
previous schemes. Initially the barrier is ramped up to the higher
value ($V_0=1450$~Hz) within 300~ms, then the barrier is kept
constant for 100~ms and then the barrier is ramped down to the lower
value ($V_0=1140$~Hz) within 300~ms. The fluctuations measured
according to scheme 3 are smaller than those measured according to
scheme 2, indicating a thermalization process. This leads to the
counterintuitive behavior that phase fluctuations can be decreased
by thermalization processes.

To get a more quantitative handle on thermalization we compare the
coherence factor measurements with the theoretical prediction from
Eq.~\ref{eq.classical}. We introduce an effective tunneling coupling
$E_J^{\rm eff}$ to account for out of equilibrium situations and
deduce $k_BT/E_J^{\rm eff}$ from the experimental data shown in
Fig.~\ref{fig.coherence}. In Fig.~\ref{fig.thermalization} (b) the
ratio $E_J^{\rm eff}/E_J$ is shown as a function of the tunneling
time $\tau_p=2\pi/\omega_p \propto 1/\sqrt{E_J}$. We find that for
the chosen ramping time of 300~ms the effective tunneling coupling
is only equal to the numerically calculated tunneling coupling for
$\tau_p<50$~ms. Thus for the application as a primary thermometer
care has to be taken to choose the appropriate tunneling parameters.
The observed increase of $E_{J}^{eff}$ for large tunneling times
($\tau_p>50$~ms which corresponds to $E_J<60$~nK), could be
explained by the fact that the system may still have not reached
equilibrium after the 300~ms ramp.

\section{Application of the thermometer to heat capacity measurement} \label{sec_thermometry}
From the previous discussion it follows that if the tunneling rate
is sufficient to guarantee thermal equilibrium ($\tau_p<50$~ms for a
300~ms ramping up time of the potential barrier), the measurement of
the phase fluctuations and an independent determination of the
coupling strength $E_J$ constitute a primary thermometer.

\begin{figure}[h!]
\centering\includegraphics[totalheight=6.5cm]{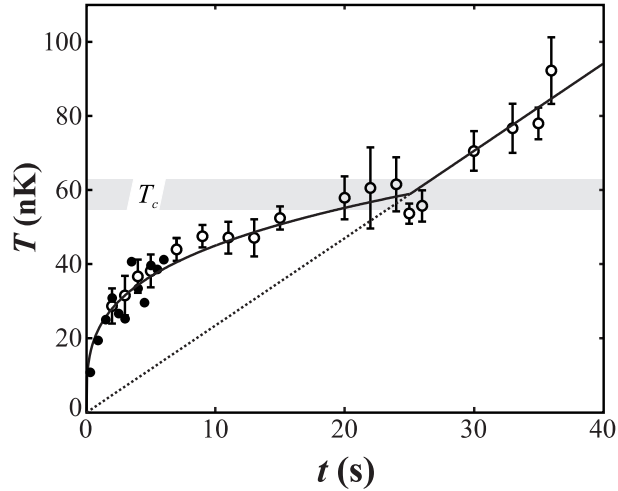}
\caption{\label{fig.thermometry} Thermometry using phase
fluctuations measurements. The graph shows the application of the
phase thermometer. The temperature of a degenerate Bose gas is
plotted as a function of the holding time in the harmonic trap. The
blue points correspond to measurements using the phase fluctuation
method (for the few points where $\tau_p>50$~ms we slightly correct
the calculated temperatures using the effective Josephson energy
plotted in Fig.~\ref{fig.thermalization} (b)) and the open circles
to measurements applying the standard time of flight method. In the
overlap region both methods lead to the same results showing the
applicability of the noise thermometer. The black line corresponds
to a fitting function assuming a constant transfer rate of energy, a
power law for the temperature dependent heat capacity below the
critical temperature and a constant heat capacity above. The
measurements demonstrate the deviation of the heat capacity of the
Bose gas from the classical gas.}
\end{figure}

To test the applicability of the new thermometer we measure the
heating up of a degenerate Bose gas in a 3D harmonic trap. For this
the BEC is prepared at the lowest accessible temperature and the
cooling is turned off. The phase fluctuations are then measured
after different holding times by ramping up the barrier within
300~ms to barrier heights corresponding to a coupling strength on
the order of the thermal energy scale. About 60 interference
patterns are produced for every holding time in the harmonic trap.
For longer holding times, where the thermal fraction becomes
visible, also time-of-flight measurements from the 3D harmonic trap
are performed. The results are shown in Fig.~\ref{fig.thermometry}.
The blue points correspond to temperature measurements using the
phase fluctuation method and the open circles to measurements using
the standard time of flight method. We observe the heating after a
holding time of up to 36~s. After 25~s the critical temperature of
$T_c=59.1$~nK is reached and the condensate fraction vanishes. For
these measurements the total number of atoms in the trap is kept
constant, thus the phase fluctuation measurements can only be
performed up to 6~s as long as the condensate fraction is large
enough to observe clear interference patterns. Below 30~nK the time
of flight method cannot be applied as the fraction of thermal atoms
in this regime is too small and cannot be fitted.

The increase of the temperature with time can be explained by taking
the dependence of the heat capacity on temperature of a Bose gas
into account and assuming a constant transfer of energy. For short
holding times a fast increase of the temperature corresponding to a
large heating rate can be observed. The heating rate decreases
continuously until the critical temperature is reached, above the
critical temperature the heating rate stays constant. We attribute
this decrease of the heating rate to an increase in the specific
heat as expected for a degenerate Bose gas. We assume the specific
heat to be proportional to $(T/T_c)^d$ below $T_c$ and constant
afterwards \cite{DeGroot:01}. The expected evolution of the
temperature with time $t$ is then
\begin{equation}
T(t) = \cases{\sqrt[d+1]{h_0  T_c^d t + T(0)^{d+1}} &for $T_{\rm fit}<T_c$\\
h_0  t + T(0) &for $T_{\rm fit} \ge T_c$ \, , \\}
\end{equation}
where $h_0$ is the constant energy transfer rate and the critical
temperature is deduced from independent measurements of the trap
parameters and atom numbers $T_c  \approx  0.94 \cdot \hbar/k_B
\cdot (\omega_x\omega_y\omega_z)^{1/3} \cdot N^{1/3} = 59.1
\textrm{~nK}$. Using this function to fit the observed temperature
increase, we obtain $h_0 = 2.4(1)$~nK and a dimensionality parameter
$d=2.4(4)$.

The most likely source of heating in these experiments are
fluctuations of the trap position and the trapping frequencies. The
heating due to fluctuations of the trap position corresponds to a
constant increase of energy per time and particle and the heating
due to fluctuations of the trapping frequencies (parametric heating)
to an exponential increase of the energy \cite{Savard:97}. The
fitting with a function taking both heating processes into account
reveals that the additional increase due to parametric heating is
very small and results in a correction of the temperature of below
$7\%$ after the 36~s with $d=2.7(7)$. Thus, the assumption of a
constant transfer rate of energy describes the experimental
situation very well.

The observation of the heating for low temperatures represent the
extension of the heat capacity measurements already performed in the
early days of BEC \cite{Ensher:96} to the low temperature limit. The
data clearly revealed that the heat capacity of a degenerate
interacting Bose gas is smaller than the classical gas prediction
for temperatures below $0.7 \times T_c$. The dimensionality deduced
from our data is slightly smaller than the theoretical prediction
$d=3$ for the heat capacity of an ideal Bose gas confined in a 3D
harmonic trap, as expected due to the presence of atom-atom
interaction \cite{Minguzzi:96}. Clearly the dependence of the heat
capacity on the temperature with $d>1$ confirms the prediction of
the third law of thermodynamics \cite{Feynman:63} stating that the
heat capacity has to vanish in the zero temperature limit.

From this analysis we can conclude that the phase fluctuation
measurements can be applied for thermometry without the need for
calibrating the thermometer with independent methods. However, a
more detailed understanding on the thermalization process and the
relevant timescale is necessary to be able to predict the range of
validity for the measurements.

\section{Conclusion}
In summary we have presented a detailed analysis of a new method for
measuring ultralow temperatures of degenerate Bose gases in a regime
where standard time of flight methods cannot be applied. These
temperature measurements were done by investigating thermally
induced fluctuations of the relative phase between two weakly
coupled Bose-Einstein condensates. We have compared the
experimentally obtained coherence factor with the theoretical
prediction using a standard two mode model at finite temperature. We
found quantitative agreement over a wide range of the relevant
scaling parameter. With this a primary thermometer is realized.
However, it is important to note that due to the approximations in
the theoretical model this method leads to good results for low
temperatures (far below the critical temperature) and if quantum
mechanical fluctuations are negligible. Furthermore, care has to be
taken for the preparation of the BJJ such that thermal equilibrium
is guaranteed, i.e. the ramping of the barrier has to be much slower
than the tunneling time.

The application of this noise thermometer was demonstrated by
measuring the heating up of a quantum degenerate Bose gas. The
observed temperature increase reveals in a direct way that the heat
capacity of the Bose gas below the critical temperature deviates
strongly from the heat capacity of a classical gas and vanishes in
the zero temperature limit as predicted by the third law of
thermodynamics.



\ack We thank T.~Bergeman very much for the numerical calculation of
the relevant parameters and the valuable theoretical support. We
would also like to thank A.~Trombettoni for discussions, M.~Albiez
and J.~Foelling for their contributions to the experiments,
M.~Scherer for implementing the 3D Gross-Pitaevskii code, and
D.~Weiskat for his indispensable help with the electronics. This
work was funded by Deutsche Forschungsgemeinschaft
Schwerpunktsprogramm SPP1116 and by Landesstiftung
Baden-W\"urttemberg - Atomoptik. R.~G.~thanks the
Landesgraduiertenf\"orderung Baden-W\"urttemberg for the financial
support.

\section*{References}
\bibliographystyle{unsrt}
\bibliography{library}

\end{document}